# Sustainable Edge Computing: Challenges and Future Directions

Patricia Arroba[a], Rajkumar Buyya[b], Román Cárdenas[a], José L. Risco-Martín[c], José M. Moya[a]

a       Laboratorio de Sistemas Integrados (LSI)
        CCS–Center for Computational Simulation
        ETSI Telecomunicación
        Universidad Politécnica de Madrid, Spain

b       Cloud Computing and Distributed Systems (CLOUDS) Lab
        School of Computing and Information Systems
        The University of Melbourne, Australia

c       Department of Computer Architecture and Automation
        Universidad Complutense de Madrid, Spain

## Abstract

An increasing amount of data is being injected into the network from IoT (Internet of Things) applications. Many of these applications, developed to improve society's quality of life, are latency-critical and inject large amounts of data into the network. These requirements of IoT applications trigger the emergence of Edge computing paradigm. Currently, data centers are responsible for a global energy use between 2% and 3%. However, this trend is difficult to maintain, as bringing computing infrastructures closer to the edge of the network comes with its own set of challenges for energy efficiency. In this paper, we propose our approach for the sustainability of future computing infrastructures to provide (i) an energy-efficient and economically viable deployment, (ii) a fault-tolerant automated operation, and (iii) a collaborative resource management to improve resource efficiency. We identify the main limitations of applying Cloud-based approaches close to the data sources and present the research challenges to Edge sustainability arising from these constraints. We propose two-phase immersion cooling, formal modeling, machine learning, and energy-centric federated management as Edge-enabling technologies. We present our early results towards the sustainability of an Edge infrastructure to demonstrate the benefits of our approach for future computing environments and deployments.

## I. Introduction

The Edge computing paradigm originates as a solution to increasing data rates due to data-demanding applications. A growing amount of data is being generated from Internet of Things (IoT) applications, deployed to improve the efficiency and effectiveness of various sectors and systems, such as healthcare, transportation, energy, and agriculture. This massive injection of data into the network can lead to its saturation, causing significant delays. However, this is a critical problem, as many IoT applications have strict latency requirements to ensure timely and accurate responses as they involve real-time data processing and decision-making. For example, IoT-enabled devices can monitor a patient's health remotely and alert healthcare professionals if there are any abnormalities, leading to faster and more accurate treatment. Edge computing helps solve this issue by bringing computing resources closer to the data sources, reducing the amount of data that needs to be sent to the Cloud for processing. As data is processed and analyzed on a more local basis, the benefits are numerous: (i) user-perceived delays are reduced and speed is increased, (ii) network costs are reduced and less bandwidth is used, and (iii) local reliability can be provided in case of connectivity problems [1].

Based on Gartner's predictions regarding Edge computing [2], more than 50% of the data managed by enterprises will be generated and processed outside the Cloud by 2025. According to Cisco, Edge computing technology ranges from Multi-access Edge Computing (MEC), cloudlets and micro data centers, to fog computing (between the Cloud and Edge devices) [3]. But, apart from the theoretical concept, how are Cloud providers reaching the Edge today? The answer to this question is that hyperscale operators are now everywhere. Cloud providers that own massive Clouds, which in most cases are found in remote areas, are moving toward the users in many ways. However, in the first place, they are adopting a less risky approach, which is to use local DCs, as colocation DCs, to host their Edge nodes (e.g., IT equipment, from small racks to entire data rooms). So, the boundaries of Cloud, hybrid-Cloud, and Edge computing are blurring somehow.

## A. Sustainability Impact of Moving Toward Edge

For leading Cloud providers such as Google, Amazon, or Microsoft [4], we find very sustainable and efficient Cloud DCs owned and operated by them. Today, they claim to be carbon neutral, and they aim to be Net Zero Carbon by 2030-2040 (i.e., removing as much carbon as they emit) [5]. On the other hand, they are also very efficient in using the energy they consume. To evaluate this performance, we can rely on the Power Usage Effectiveness (PUE), which is a ratio that describes how efficiently the energy is used in a data center (total facility energy divided by the energy used by the IT devices). So, this metric allows us to understand the energy overhead, mostly used for cooling down the IT resources, for every unit dedicated to computing. According to the Uptime Institute, in 2022, the average PUE of the data center industry was around 1.55 [6], revealing that approximately 35% of the total energy budget is consumed by the cooling systems.

Google, Amazon, and Microsoft claim a PUE around 1.1 [7] [8] [9], which means that their energy overhead is six times lower than the industry average. But how do they manage to be so efficient? For instance, if we examine the locations of the DCs owned and operated by Google in 2022, as seen in Figure 1.a, we find that most of them are in: (i) cold areas, where cooling is much more efficient so PUEs would be better, and (ii) areas that support renewable energy production, thus reducing the carbon footprint. So, they use cleaner energy, and they use it more efficiently. If we examine the latest data centers built by these Cloud providers, we can see they have several things in common. They occupy a large area, necessary not only to host the data rooms but also for renewable energy production (windmills and solar panels) and the cooling infrastructure, and they require a large financial investment. So, whether the Cloud is in an optimized, cool, and spacious location, and if it is adequately funded, it can be very sustainable and efficient. But what happens when the IT infrastructure needs to be closer to the data sources?

Figure 1.b represents Google's edge nodes (more than 1,300) closest to the users. Edge nodes are Cloud providers' servers supplied inside the network of local operators and internet service providers. Many are placed in colocation DCs, from racks to entire data rooms in over 200 countries and territories. Edge nodes are Cloud providers' servers supplied inside the network of local operators and internet service providers, and many of them are placed in colocation DCs, from small racks to entire data rooms. In 2020, the colocation market grew by 22% and by 9%, for wholesale and retail colocation respectively, due to these hyperscale operators' revenues [10]. These edge nodes provide services that are popular with the local host's user base, reducing latencies and network saturation.

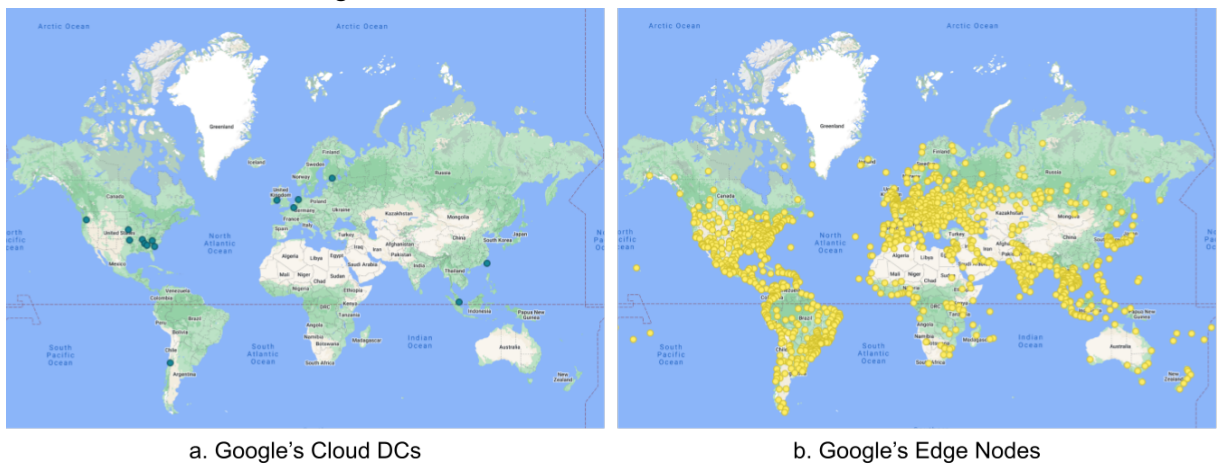

a. Google's Cloud DCs    b. Google's Edge Nodes

*Figure 1: Google's Cloud data centers and Edge nodes.*
[https://peering.google.com/#/infrastructure]

But, as the average PUE worldwide is about 1.55, we can see that the energy efficiency when IT is moved closer to the edge drops significantly. Although local DC operators have implemented numerous energy efficiency policies, reaching hyperscalers' PUEs is impossible in most cases due to economic, technical, or geographical constraints [6], causing the average PUE to stall, as seen in Figure 2. In addition, as air cooling remains the dominant technology even in new data centers, new processors with higher thermal power will penalize cooling efficiency. As a result, the industry average PUE is expected to rise over the next few years before continuing to decline. So, the further we move toward the edge, the more barriers to sustainability and energy efficiency we will find.

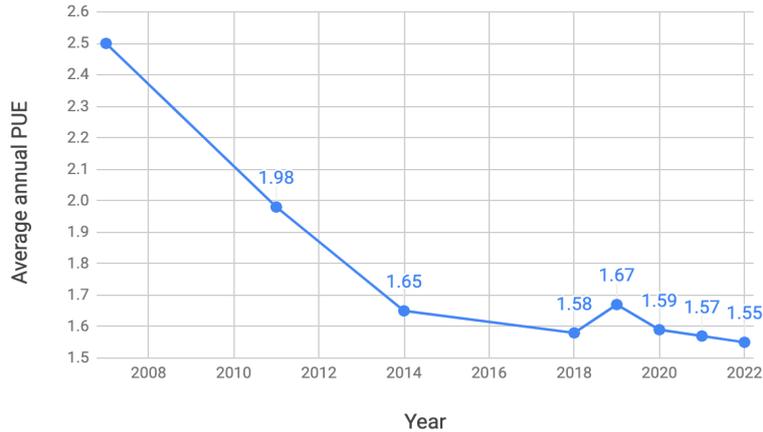

*Figure 2: Evolution of PUE in the data center industry.*

In this context, this paper presents our early steps toward the sustainability of future Edge computing infrastructures. **Our efforts are directed at achieving** (i) an energy-efficient and economically viable deployment, (ii) a fault-tolerant automated operation, and (iii) a collaborative resource management to improve resource efficiency.

The remainder of this paper is organized as follows. In Section 2, we identify the main limitations of applying Cloud-based approaches close to the data generation sources. We present the research challenges to Edge sustainability that arise from these constraints and propose an innovative approach to address them. Then, Section 3 to 5 introduce 2-phase immersion cooling, formal modeling, machine learning, and energy-centric federated management as Edge-enabling technologies. To demonstrate the potential benefit of our approach combining these technologies, Sections 6 and 7 present different use cases. In Section 8, the main conclusions and future directions are drawn.

## II. Open Challenges and Envisioned Approach

According to Gartner [2], the number of Internet of Things (IoT) devices is expected to triple from 2020 to 2030, leading to a growth in data injected into the network. The fastest growing areas will be manufacturing and natural resources, with a Compound Annual Growth Rate (CAGR) of 18%, healthcare providers (13% CAGR), smart buildings (12% CAGR), and automotive (10% CAGR). Based on these trends, the explosion of new data-intensive applications in the next few years will again saturate existing facilities. Therefore, the Edge is expected to increase its computing density by deploying new infrastructure in the near future.

Therefore, three factors must be considered in this context. First, this new Edge infrastructure will be closer to the users. Second, the closer we get to the edge, the lower the energy efficiency because the location conditions are potentially worse. Third, as Edge computing will enable the digitization of society, we should aim for a deployment that meets the needs of future applications.

Driver assistance systems, personalized medicine, and online gaming are just a few examples of applications that Edge computing will facilitate and enhance. These applications need intensive computing, ready for massive data processing through machine learning. And this must be done close to the users because the data rates will be massive, and the delays critical. In this scenario, it makes sense to envision future Edge deployments as small, compute-intensive Edge Data Centers (EDCs) located closer to the data sources.

## A. Limitations of Current Approaches and Key Open Challenges

In this context, Edge computing is much more disruptive than we expected, not an incremental solution from the Cloud. The current technology that has enabled a very efficient development for Cloud computing has several limitations when we apply it to the deployment, operation, and management at the edge of the network.

**Issues related to infrastructure deployment at the edge:**
As we have discussed previously, the closer we get to the edge of the network, the less energy-efficient the infrastructures become (mainly for cooling), with a significant increase in the average PUE from 1.1 to 1.55. In addition, the most efficient data centers occupy large areas that, in most cases, will not be available when we bring computing closer to users, as they are mainly located in urban environments.

We also must deal with energy efficiency and size in non-optimal locations while reducing costs, as numerous EDCs will be deployed. However, technological solutions needed to improve efficiency and reduce the physical footprint are usually expensive. For all these reasons, the first key challenge we need to address is how to provide an environmentally and economically sustainable deployment for Edge computing.

**Issues related to distributed dynamic operation:**
Cloud operation is carried out on-site, as these large, centralized infrastructures have personnel dedicated to the maintenance and decision-making arising from the day-to-day performance of the data center. On the contrary, EDCs will form a large and highly distributed computing infrastructure where local operation becomes impractical.

This makes learning-centric automatic operation one of the major research questions for dynamic distributed infrastructures. But what happens when anomalous situations beyond our control arise (e.g., security attacks, hot spots in data rooms, system malfunctions, service outages, etc.)? In the Cloud, these anomalies are handled by operators, but if the EDCs are unattended, we need tools to avoid irreparable equipment damage causing significant financial losses. In addition, service outages in health and safety-critical applications (i.e., personalized medicine and driving assistance) expose the physical integrity of users to risk. Therefore, the second key challenge to be addressed is how to enable fault-tolerant automatic operation for future Edge computing.

**Issues related to the efficient use of available resources:**
Large hyperscale operators' Clouds aggregate user requests from an entire region. Therefore, the demand profiles arriving at the data center are smoother as they are not significantly affected by user mobility. In Edge computing, user demand is processed locally, and mobility will result in handovers between small data centers in the same region (e.g., transferring workloads at the Radio Access Network (RAN) level or between RANs across the backbone). This is why the processing demand perceived by each of the EDCs will be drastically influenced by the mobility of the data generation sources, presenting a very high variability (i.e., daily or seasonal variation patterns).

In practice, it is unfeasible to scale EDCs for their peak demand because the infrastructure would be underutilized and extremely expensive. But on the other hand, if Edge supports critical applications, we cannot compromise the quality of service perceived by end users. For these reasons, the third key challenge to ensure the sustainability of Edge computing is to deliver effective collaborative resource management.

### B. Envisioned Approach

Sustainable deployment, fault-tolerant automatic operation, and collaborative resource management are the three key challenges to enable a sustainable Edge. These open research questions need to be addressed as the future development of a competitive Edge infrastructure will depend on its economic and energy sustainability. In this research, we propose to address them through three different approaches, which combined, aim to deliver the dense, compute-intensive Edge required to commoditize the digitization of society. Figure 3 shows the high-level architectural framework that drives our approach.

We propose to improve the environmental and economic impact of deployments by incorporating novel cooling systems, such as two-phase immersion cooling. The reduction in consumption, physical footprint, and costs offered by this technology, regardless of the environmental conditions (e.g., temperature and humidity), makes it particularly interesting for improving sustainability at the edge of the network.

The robustness, modularity and integrability of formal modeling in a Model-Based System Engineering (MBSE) framework can significantly improve decision-making in a complex environment such as a data center. We, therefore, aim to combine mathematical modeling with machine learning algorithms in a learning-centric approach to provide fault tolerance to automatic operation of a distributed nature, which is inherent to Edge computing.

The diversity of computing, cooling and renewable energy generation systems, infrastructure locations, and fluctuating user demand, energy prices, and weather, among others, have an impact on the energy usage and cost of Edge computing. Hence, we propose to leverage this heterogeneity through Edge federations, enabling collaborative management to dynamically improve resource utilization efficiency.

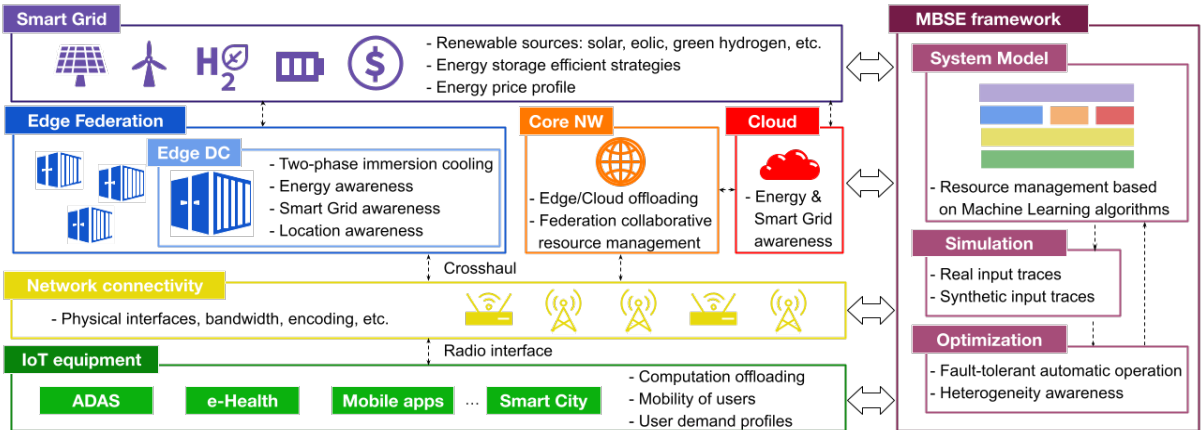

*Figure 3: A distributed environment for IoT applications integrating two-phase immersion cooling and a Modeling, Simulation, and Optimization framework for Edge federations.*

## III. Two-phase Immersion Cooling for Sustainable Edge Deployments

Two-phase immersion cooling is emerging as a technology capable of addressing many of these challenges at the edge and improving the sustainability of existing data centers. In two-phase immersion cooling, the IT is immersed in a sealed tank filled with an engineered fluid that evaporates and condenses cyclically at very high operating temperatures (above 60ºC).

When the fluid is in contact with very hot surfaces (e.g., CPU or GPU chips), the fluid evaporates. The condenser (which recirculates glycol water in a secondary circuit to exchange heat with the outside) then condenses this vapor back to the liquid phase, as shown in Figure 4. These phase changes are very efficient in extracting heat from the system [11], resulting in a reduction in cooling consumption of up to 95% compared to traditional methods.

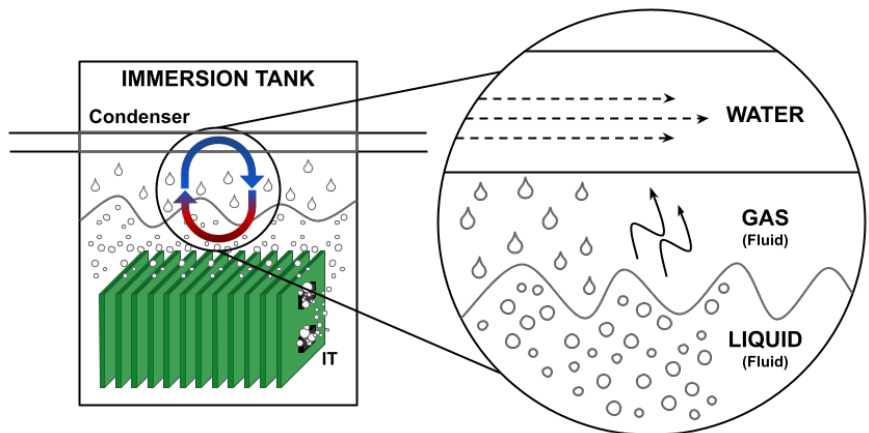

*Figure 4: 2-phase immersion cooling.*

### A. Benefits of 2-phase immersion for Edge

**Better energy efficiency and weather flexibility:**
The change between gas and liquid phases of this technology is very efficient in extracting heat from immersed systems and occurs at elevated temperatures above 60°C. This results in a significant reduction in cooling consumption of up to 95%, achieving PUEs of 1.02-1.03, not seen before but in free cooling. In addition, it has a low global warming potential (below 1), thus complying with restrictive environmental regulations.

One of the main advantages of two-phase immersion for Edge computing is that it provides these benefits regardless of the location of the infrastructure and the outside temperature. As the operating temperatures of this technology are above 60°C, the setpoint temperatures of the cooling system can be very high. This can provide a PUE of around 1.03, significantly below the industry average (PUE=1.55) and better than the average for hyperscale operators (PUE=1.1), even in edge locations with hot climates.

**Higher power densities and reduced physical footprint:**
Power density in data centers is a key metric for designing computing and cooling infrastructures [6]. In recent years there has not been a significant increase in rack power, and the industry continues to work with typical 4-6 kW cabinets. One of the main reasons for this deadlock is air cooling, still the most widely used in the industry, which is limited by the low heat transfer coefficient of air. Moreover, this traditional technology requires hot and cold aisle configurations and a large amount of equipment (e.g., chillers, computer room air conditioner and handler units, etc.), which significantly penalizes its physical footprint.

Two-phase immersion cooling enables a 60-fold increase in the power density in a rack, given its high heat transfer capacity, allowing up to 250kW per rack [11]. It also significantly improves the physical footprint of the infrastructure. This technology shrinks the floor space of data centers by up to 10 times, providing more computing power in less area, from 10kW/m2 with traditional cooling to 100kW/m2 with 2-phase technology. This allows EDCs to be downsized to fit in urban environments close to users, ensuring a very high computing capacity to support applications with demanding requirements.

**Low cost:**
The heat extraction performance and efficiency of two-phase immersion cooling significantly reduce infrastructure costs. As we minimize cooling consumption by up to 95%, we considerably improve the energy budget (only 1% energy overhead). In addition, we no longer need as much cooling equipment (also cutting water usage) or as much floor space as with traditional cooling, so capital expenditure will also be significantly reduced. This cost reduction is key for Edge due to the large volume of EDCs needed to meet application demand locally.

# IV. Model-Based Systems Engineering and Machine Learning for Edge Automatic Operation

Today's data centers already integrate elements of artificial intelligence to optimize the management of computing and cooling systems. The systems monitor the environment and inject the collected data into algorithms that learn to optimize performance decisions. According to the most recent survey of the Uptime Institute [6], more than half of data center owners and operators would rely on appropriately trained machine learning models to make operational decisions. Automated operation is a choice for DCs managed on-site by operators, but given the distributed nature of Edge computing, the closer we bring IT to the users, the more difficult it will be to have staff deployed in the infrastructures. Therefore, in Edge computing, the automation of the operation becomes essential, and standardization of the environment must be an integral part of the operational model to control and manage Edge systems [1].

In this context, we aim to combine Model-Based Systems Engineering and Systems Optimization through Simulation with formal specification and machine learning to provide a formal Edge computing model with two objectives. The first is to standardize the design and operation of the infrastructure using mathematical formalisms. The second objective is to help optimization algorithms explore the system behavior and make improved decisions automatically.

## A. Formal Modeling for the Operation Digitalization

In this paper, we propose a novel approach that integrates Modeling, Simulation, and Optimization (M&S&O) methodologies in the Systems Engineering (SE) technical process for developing complex systems. Our Modeling, Simulation and Optimization-Based Systems Engineering (MSOBSE) strategy emphasizes using formal modeling tools for building more robust and reliable systems. Simulation is the primary tool to verify that the proposed model meets the system requirements. In addition, it embeds optimization techniques to automate design decisions that improve the Key Performance Indicators (KPIs) compliance of the system under development. Figure 5 shows the proposed MSOBSE workflow.

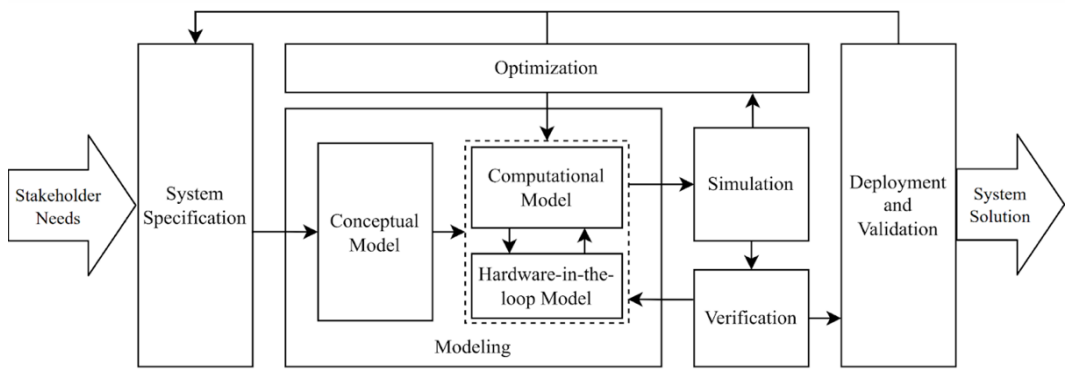

*Figure 5: Workflow of the proposed MSOBSE technical process.*

As in previous SE technical processes, we must first study the needs of all system stakeholders (e.g., customers, contractors, and developers). As a result of this study, we will define the KPIs that make up the system requirements. Once the target system is specified, we start with the system modeling process. First, we provide a conceptual model of the solution. Then, the conceptual model is translated into a computational model. Computational models are computer programs that simulate and analyze the system behavior using a virtual representation instead of a physical prototype. We can run simulations over the computational model to verify that the model behaves as specified by the conceptual model. Usually, this process follows an incremental approach: the conceptual model is broken down into multiple submodels. Each subcomponent is translated into a computational model, simulated, and verified independently. As submodels are proven, we merge them and verify the resulting model. Eventually, we will have a computational model that satisfactorily captures the behavior of the formal model of the system.

Complex systems, such as data centers, have numerous degrees of freedom. In these situations, an optimal design of all the elements in the scenario is not trivial. To tackle this issue, we propose integrating automated optimization algorithms in the system development process. These algorithms will adjust the verified computational model to obtain the solution that satisfies the system specifications without incurring high development costs. If we cannot get a design that meets all the system requirements, the specification will be revised, and we will repeat the modeling, simulation, and verification process. After optimizing the computational model, we start the system implementation process. Hardware-In-The-Loop (HIL) models gradually replace parts of the computational model with an actual implementation of the modeled system components [12]. Then, we test and validate the implemented elements by running real-time simulations over the HIL model. If the component does not behave as its corresponding model, we will review the conceptual model to fix potential mistakes. Eventually, we will end up with a complete implementation of the system under study. The developed solution is the product of a validated conceptual model in which components are optimized for meeting all the system specifications while reducing development expenses.

### B. ML for Fault-tolerant Decision Making

Machine learning-based solutions are now a reality in data centers. Machine learning currently provides very accurate dynamic behavioral models of infrastructure elements (e.g., temperature, humidity, power consumption, etc.). It also helps operators to optimize control decisions (e.g., temperature setpoints, workload allocation, and scheduling). To properly train machine learning models, we only need a large amount of data collected by a monitoring system. This works very well for the standard behavior of real systems in production. But what happens when anomalous situations occur (i.e., equipment or monitoring system failures)?

Failures in data centers can lead to outages with severe economic consequences. In the last three years, around 60% of the operators surveyed [6] suffered an outage due to problems related mainly to power (37%), computer systems (22%), and cooling (13%). The last outage they experienced had cost between $100,000 and $1 million for 45% of these data centers and more than $1 million for another 25% of them. Most operators agree that these outages could have been avoided through better procedures, management, configuration, and training. And this problem would be even more severe for distributed, unattended EDCs, with potentially longer downtimes and, thus, higher costs.

We aim to improve fault-tolerant management in these infrastructures to make decisions while predicting, avoiding, or automatically mitigating outages and equipment failure. As these situations are rare, only a tiny percentage of the collected data presents anomalies. Therefore, the datasets with which machine learning algorithms are trained are highly biased, and management systems cannot learn to make decisions when failures occur. Reproducing an anomaly in an actual data center is potentially dangerous for the equipment and may consume a lot of time and energy. In this paper, we propose to use machine learning to optimize the system management and provide balanced datasets by generating realistic synthetic data that also reproduce anomalous situations. This is a major challenge due to the complexity of multivariate data center scenarios, as for synthetic datasets to be realistic, they must maintain nonlinear correlations between their variables when an on-demand anomaly is generated.

## C. Benefits of integrated MSOBSE and ML for Edge

Edge computing requires an infrastructure that is currently in the design and deployment phase and hence benefits from an M&S&O standardization framework to explore its sustainable development. Our proposed approach also enables the optimization of its dynamic operation by incorporating machine learning for decision-making in anomalous situations. This provides the fault-tolerant automation required for the decentralized and unattended management of EDCs. The main benefits of our approach are as follows.

**Formal foundation:**
A modeling approach that relies on a mathematical formalism provides greater completeness and robustness to the system model. Therefore, it facilitates the timely detection of errors in the conceptual model. This strategy allows us to develop the system earlier and at a lower cost.

**Model modularity:**
As previously explained, modeling a complex system is an iterative process. System components are modeled and verified separately. Then, we progressively merge the verified models and prove the resulting model again. A modular modeling technique is an excellent choice in these scenarios since it simplifies the integration and verification process of the Edge infrastructure with preexistent technology (e.g., Clouds, IoT devices, and network elements).

**Hardware-In-The-Loop integration:**
The proposed workflow integrates HIL modeling techniques in the system implementation and validation phases. We aim for a suitable modeling framework that eases the transition from a computational model to a HIL model without imposing additional development times. So, any computational models could be replaced by real hardware allowing the MSOBSE framework to perform in pure simulation contexts, real scenarios, or hybrid environments. Consequently, the framework can explore the entire system's behavior, being aware of its impact on both formal models and actual equipment (e.g., resource usage, energy consumption, and QoS perceived by the users).

**Fault-tolerant Decision Making:**
Finally, the MSOBSE workflow must provide a complete interface for optimizing the system under development. As shown in Figure 6, ML can be integrated with the MSOBSE framework to produce unbiased/biased training environments. The ML-based synthetic data generator allows the MSOBSE framework to train with sufficient data from standard system behavior and anomalies. This way, the resource management optimization algorithms will learn to explore decision-making for system failure scenarios based on the impact observed in the formal model. Our proposed approach aims to provide trained ML-based models that enable fault-tolerant decision-making in real Edge infrastructures.

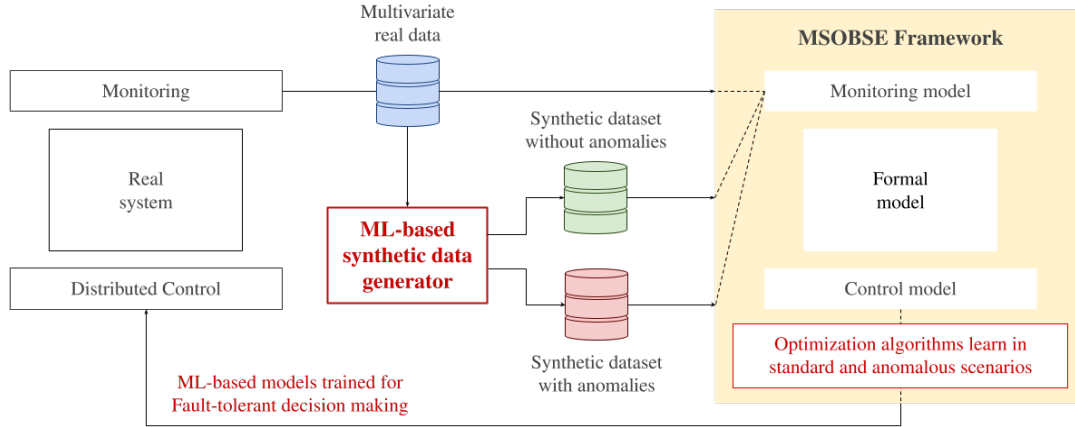

*Figure 6: Workflow of the integrated approach for fault-tolerant decision making.*

# V. Edge Federations for Collaborative Resource Management

The profile of service requests perceived locally at the network's edge is significantly variable due to user mobility. Dimensioning EDCs for peak demand is impractical as it is expensive and resource usage is inefficient. However, service availability must be ensured, given the criticality of Edge target applications. Thus, collaboration is necessary for the Edge to absorb peak demand while guaranteeing the quality of service. In this paper, we propose the federated management of EDCs to improve energy and cost sustainability and maintain the quality of service perceived by users.

### A. Harnessing Heterogeneity for Increasing Energy Efficiency

Heterogeneity is an intrinsic challenge faced by federated management. As shown in Figure 7, the industry considers new cooling systems and renewable energies as the main drivers of sustainability [6]. The gradual adoption of emerging cooling technologies (e.g., liquid cooling, single-phase, and two-phase immersion) and green energy sources (e.g., solar photovoltaics, wind power, or green hydrogen) will increase the physical heterogeneity of data centers. Moreover, the geographical distribution of EDCs, closer to users, is another source of heterogeneity for federated management. Weather, energy prices, smart grid capabilities, local user demand, and their fluctuation profiles will depend on the location of each Edge infrastructure.

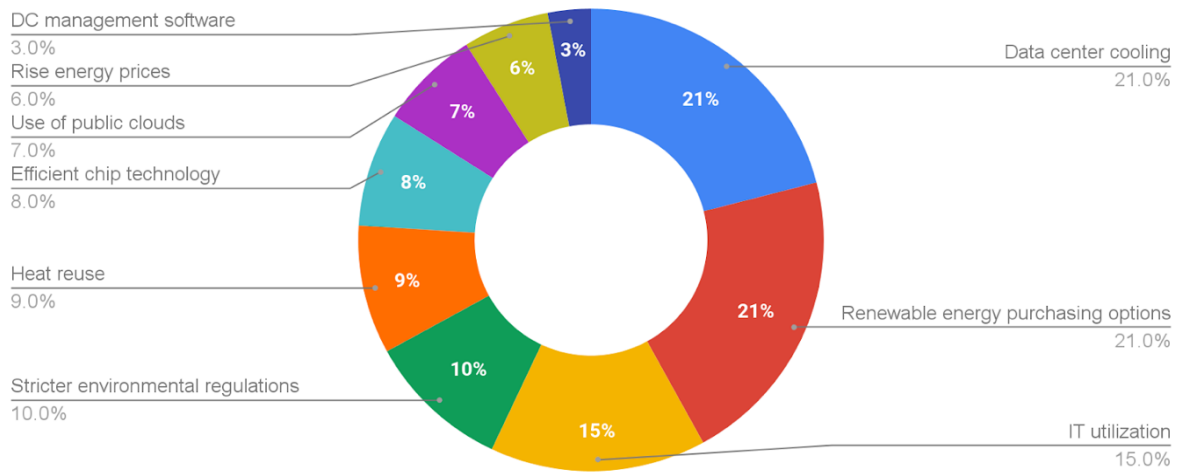

*Figure 7: Drivers for sustainability gains in the data center industry.*

This heterogeneity offers a great opportunity to improve the sustainability of Edge computing through federated management. We propose to augment the resource management systems to include the energy impact of the scheduling and configuration decisions made in the federation. In this way, the federation may harness the heterogeneity of each EDC in its specific location at any given time while being aware of its status as follows.

- **Utilization of IT equipment:** dynamically driven by the assigned workload.
- **IT power consumption:** depending on IT technology (e.g., CPUs, GPUs, etc.), resource utilization, and configuration of their low-power policies.
- **Cooling power consumption:** varying with the heat generated by IT equipment, the cooling technology, and the outdoor climatic conditions (i.e., temperature and humidity).
- **Renewable energy production:** which depends on the deployed energy generation technology, and the weather (i.e., solar radiation intensity, wind intensity).
- **Energy storage:** a function of storage technology characteristics and charging and discharging policies.
- **Grid energy pricing:** set by energy companies depending on users' daily consumption patterns.
- **Quality of service requirements:** agreed specifically for the different applications and services supported by the federation.

### B. Harnessing Secondary Market Models for Improving Resource Utilization

IT utilization is the third major driver perceived by the industry for sustainability improvement, as can be seen in Figure 7. Today, there is still a lot of effort in this direction in Cloud data centers as their average utilization does not exceed 60% [13]. However, this problem can be even more severe for the network's edge infrastructure. Improving IT resource utilization is another inherent challenge of federated Edge infrastructures that support critical applications. Overprovisioning computing equipment to ensure the quality of service reduces resource utilization. This has a negative impact on energy efficiency, as even completely idle servers consume up to 70% of their maximum power [14].

In this paper, we also propose to include the secondary market model in federated Edge management, where tenant idle resources are leased to other companies to increase resource utilization. This model reduces the need for companies to deploy their Edge infrastructure by leveraging the computing resources of existing EDCs. However, this secondary market brings challenges, as workloads must coexist with applications with critical QoS and security requirements. The federation would benefit from the secondary market model by being aware of the following for each type of workload.

- **Application resource utilization profile:** e.g., CPU, memory, I/O, network, etc., and their fluctuation due to user interaction.
- **User demand profile:** which varies in different patterns such as daily, weekly, etc.
- **Contracted quality of service:** based on the nature of the application (best effort vs. critical applications).

Models that allow detailed predictions of these patterns will be key in ensuring the quality of service to users. In addition, predicting energy-related variables based on the evolution of resource demand would enable proactive management strategies. We aim to improve the sustainability and efficiency of Edge operation by leveraging heterogeneity and secondary market opportunities while maintaining the quality of service of the applications.

## VI. Use Case: Sustainable Edge Computing for Advanced Driver Assistance Systems

Advanced Driver Assistance Systems (ADAS) emerge from the need for a safer driving experience to reduce the number of accidents and casualties. The purpose of ADAS is to predict and avoid emergencies based on data collected from vehicles, their environment, and their occupants. According to Nvidia [15], this type of application generates, per vehicle, about 2 gigapixels of data per second with a processing capacity requirement of 250 trillion operations per second. Embedding IT systems in vehicles to enable this computational capacity makes the technology solutions extremely expensive, slowing down their commercial development. On the other hand, sending the data to the Cloud for processing increases service latency and could risk the physical integrity of vehicles and their occupants.

Therefore, deploying a computing infrastructure at the network's edge is essential for ADAS to manage its large data volumes and the criticality of its delays [16]. In addition, computation offloading close to the data sources will enable collaboration between vehicles and Smart Cities to improve the driving experience. Consequently, ADAS, as a precursor to the autonomous vehicle, is one of the key applications to drive the development of Edge computing, hence the importance of exploring the sustainability of its deployment and operation. This section presents our first steps towards the sustainability of an Edge infrastructure for ADAS. These steps include: (i) design of a prototype EDC with two-phase immersion cooling, exploring its thermal and power behavior, (ii) modeling the EDC prototype and simulation of its federated operation to harness cooling heterogeneity, and (iii) modeling and simulation of Smart Grid-awareness in the federation.

## A. Prototype of a two-phase Immersion Cooling EDC for ADAS

The objective of this use case is to demonstrate that our approaches can contribute to the development of a sustainable Edge infrastructure for the digitalization of society. Since 2017, we have been developing, together with the industry, a data center prototype to research the viability and benefits of two-phase immersion cooling technology for Edge computing. Our prototype, placed at the Technical School of Telecommunications Engineering at Universidad Politécnica de Madrid, Spain, is a small ship container adapted as a data room, as shown in Figure 8. The immersion tank inside the container houses a GPU computing infrastructure suited to run ML-based applications. The dry cooler placed on the top of the container is the heat exchanger with the environment. Our EDC prototype is designed for an IT infrastructure of up to 50 kW.

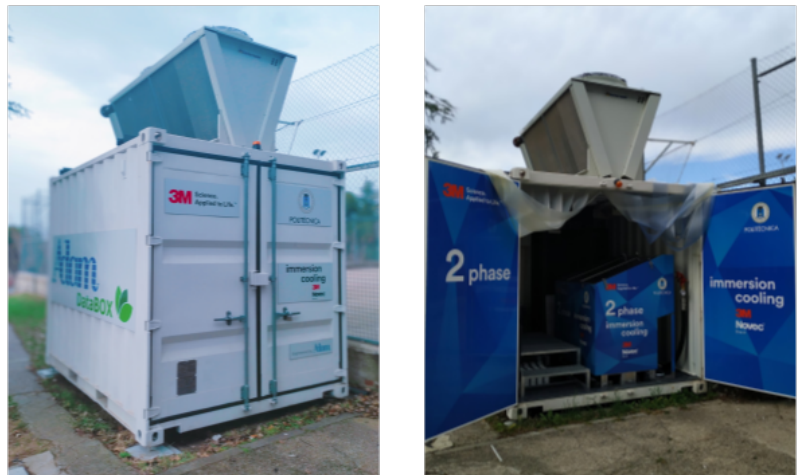

*Figure 8: Two-phase immersion cooling EDC prototype.*

Figure 9 shows the installation of the cooling system outside the container, including the expansion vessel, the drive pump, the dry cooler, and the piping necessary for its operation. In this prototype, the only contributor to cooling consumption is the pump that circulates the water from the condenser to the dry cooler in a closed circuit. Therefore, the cooling model only depends on the difference in outlet and return temperatures to the immersion tank ($T_{out}$ and $T_{in}$ respectively) and the ambient temperature ($T_{amb}$).

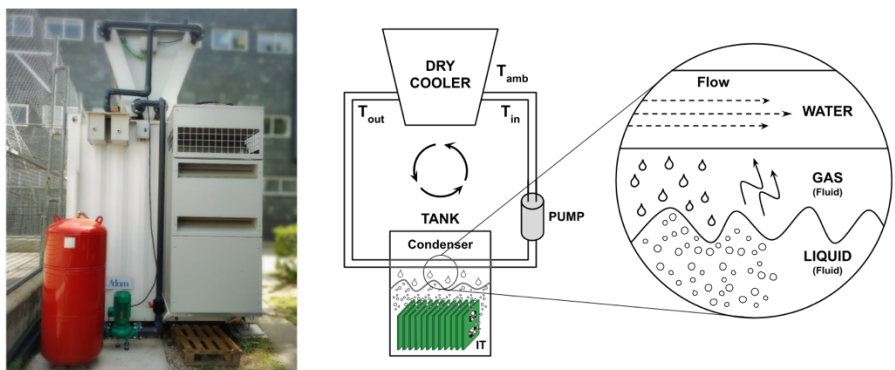

*Figure 9: Simplified two-phase immersion cooling EDC.*

According to the specifications of the Intersam model IDV-130 V-cooler used in the prototype, the following temperature ranges have been specified that interface the computing and cooling systems:

$\Delta T_1 \in [0, 20]$ (ºC)
$\Delta T_2 \in [0, 8]$ (ºC)
where:
$\Delta T_1 = T_{in} - T_{amb}$
$\Delta T_2 = T_{in} - T_{out}$

The Aero-cooler is designed to work in an air-cooled environment where the temperatures are much lower than those found in immersion cooling. Therefore, we will work at the maximum of the supported ranges (i.e., if Tamb = 32ºC, then Tin = 52ºC and Tout = 44ºC). In this scenario, we have used a Wilo IPL 50/115-0.75/2 drive pump. As the drive capacity is oversized, the cooling power will be mainly defined by the minimum flow rate it can provide, which is sufficient for the operating temperature range of the system. Therefore, the cooling power will be around 1300 W in most cases, as discussed in our previous research, which presents the Pump's power consumption model ($P_{EDC\_COOL}$) used in this work [11]. This allows us to have a theoretical PUE of about 1.03 within the state-of-the-art two-phase immersion technology. Moreover, even for a resource utilization like that found in Clouds (60%), the PUE remains around 1.04.

As for the computing infrastructure, a custom rack has been designed to immerse the GPUs (Sapphire Pulse Radeon RX 570) in the dielectric fluid. This rack allows the GPUs and their controllers to be mechanically secured. The design allows the interleaving of two racks of 9 GPUs each, increasing the computing density. It is designed to overlap several units vertically and horizontally in the immersion tank to achieve high computational densities. Figure 10.a shows a rack of 9 GPUs designed for air cooling. In our prototype, we have removed the air-cooling elements of the IT, such as fans and heatsinks, that are usually bulky, especially for GPUs. This has allowed us to fit twice as many GPUs in the same space by interlocking two racks, as shown in Figure 10.b. This configuration allows us to accommodate up to 50 kW in the container.

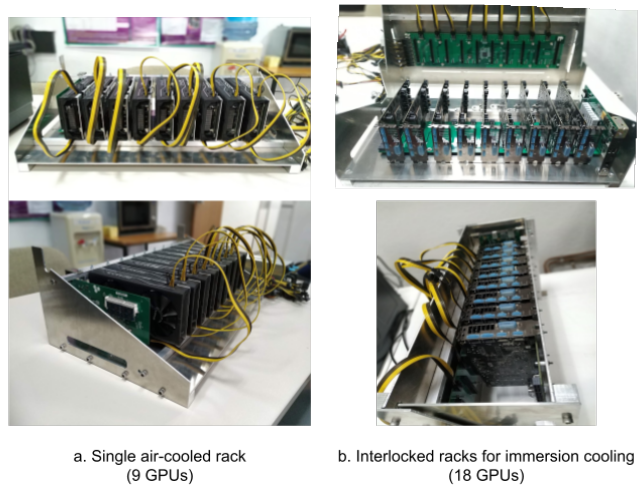

a. Single air-cooled rack (9 GPUs)  b. Interlocked racks for immersion cooling (18 GPUs)

*Figure 10: Comparison of computing density per rack.*

GPUs run Deep Learning (DL) algorithms with significantly higher performance than CPUs [17], making GPU clusters one of the best IT infrastructures for real-time video processing applications. Our ADAS application [18] is a DL-based service that alerts the user if a distraction is detected while driving. The system collects video footage from inside the vehicle and processes it through a Convolutional Neural Network (CNN) that estimates whether the driver has lost attention on the road. As DL algorithms training is computationally intensive, they are offloaded to the Edge layer, reducing the costs of the in-vehicle hardware system. This is a common practice in industry and research, which are exploring the benefits of Federated Learning to improve the efficiency of Edge computing training. Models trained at the EDC with the user's video footage are sent to the vehicle to improve the quality of predictions. As Edge infrastructure is close to the users, models can be updated more frequently with minimal delay to improve road safety.

The power profile of this workload running on the IT infrastructure of the EDC varies dynamically and depends on the number of users. In addition, two-phase immersion coolants have a very turbulent performance when the boiling point is reached. In this research, we used Novec 7100, an engineered fluid with a boiling point of 61ºC [11]. Thus, ML is an excellent alternative to obtain very accurate power models (without prior knowledge of fluid dynamics) to help us estimate the workload behavior in our EDC.

We executed the ADAS workload while modifying the main control variables (such as GPU clock frequency settings) and for different numbers of user parallel sessions, in two-phase immersion. We modeled the IT power using this data with a FeedForward Neural Network [11]. We obtained an accurate power model, $P_{IT}(t)$, as shown in Figure 11, with a normalized root-mean-square deviation (NRMSD) of about 3% and an $R^2$ of 98%.

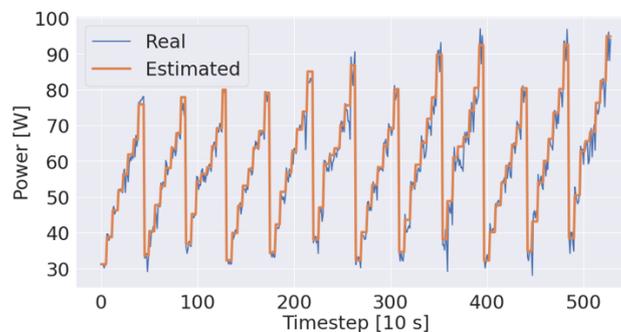

*Figure 11: Power model estimations of the immersion-cooled EDC's IT equipment.*

### B. Modeling and Simulation of Two-phase Immersion Cooling EDCs

To enable the automatic operation of EDCs as part of a federation, we propose to apply our M&S&O methodology. In this paper, we will use Mercury [19], our MSOBSE framework, to help optimization algorithms explore the system behavior and automatically improve decisions based on formal modeling. Our framework provides detailed models for the elements of the system, from vehicles to EDCs federations. We have modeled EDCs to resemble the behavior of our two-phase immersion prototype, including the obtained IT power and cooling models. Figure 12 presents the models defining the complete system and the different levels of granularity required for the EDC federation. $P_{EDC\_IT}(t)$ is the dynamic aggregated consumption of all the GPUs in the immersion tank. $P_{EDC\_COOL}(t)$ is the dynamic cooling consumption as a function of the pump's flow rate ($\phi_{EDC}(t)$) that depends on the temperature difference provided by the dry cooler ($T_{in} - T_{out}$) and $P_{EDC\_IT}(t)$ [11]. We then use the DEVS formalism [20] to define the conceptual model with which the computational models are generated [21]. Our framework simulates the computational models using the xDEVS simulation engine [22].

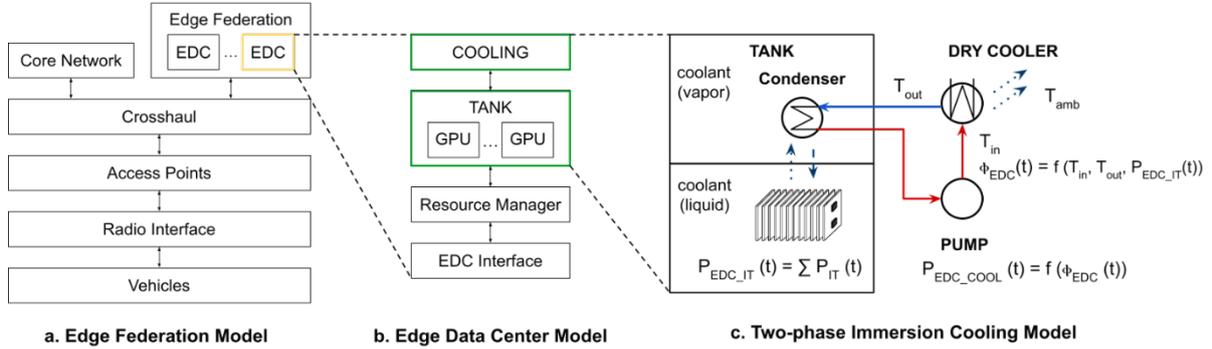

*Figure 12: Conceptual model of the system.*

**Performance Evaluation of harnessing cooling heterogeneity:**

Based on our previous work [11], we simulate EDCs operation in an Edge federation with heterogeneous cooling, as immersion will be developed progressively and in coexistence with air cooling. ADAS application requests depend on the number of vehicles in the coverage range of the access points, and their profile varies significantly with daily and weekly patterns. The EDCs of the federation collaborate in a coordinated manner to absorb this demand. We optimized the workload allocation, considering heterogeneous cooling power for the EDCs and user mobility.

We used a Reinforcement Learning actor-critic configuration to assign the workload, providing rewards to more efficient scenarios. The main optimization goal is to select the best EDC to assign new incoming sessions based on EDCs' energy status to minimize the energy consumption of the federation. On the other hand, our baseline allocated the workload to the vehicles' closest data center prioritizing the delay minimization regardless of the energy consumption. Further information on the scenario configuration, user mobility, and algorithm details can be found in our previous work [11].

To evaluate the performance of our prototype model for ADAS, we extended Mercury to include the proposed conceptual model of our two-phase immersion prototype. Our MSOBSE proposed approach helps us easily include and validate our models in the framework. Figure 13 shows the comparison of the total power consumption of the federation for the heterogeneous scenario. Our results show better efficiency for higher heterogeneity (combining air-cooled and two-phase immersion-cooled EDCs), reaching energy savings of more than 20% compared to delay-based optimization policies.

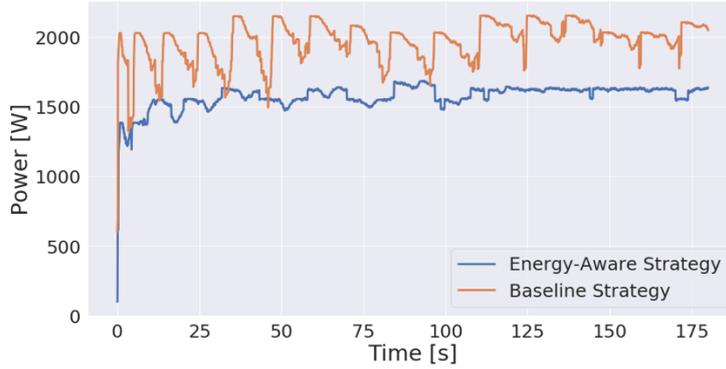
Figure 13: Power consumption comparison of the Edge federation.

## C. Modeling and Simulation of Smart Grid-awareness

We have included a Smart Grid formal model in our MSOBSE framework, which is shown in Figure 14. The energy provider (PROVR) supplies electricity to the smart grid at a variable price ($E_{price}(t)$). The solar panels produce free green power ($P_{solar}(t)$) with a variable profile depending on the solar irradiation. If this generated power exceeds the total power demand of the EDC ($P_{EDC}(t) = P_{EDC\_IT}(t) + P_{EDC\_COOL}(t)$) a surplus of energy ($P_{surplus}(t)$) is expected. Each EDC in the federation has a smart grid consumer model (CONSR) that involves its solar panels, a battery, and a controller. The storage controller makes charging and discharging decisions for the battery ($P_{charge}(t)$) based on its maximum capacity ($B_{capacity}(t)$), the energy price, and the surplus energy ($P_{surplus}(t)$). The controller also manages how much electricity is consumed from the grid ($P_{CONS}(t)$). The controller management workflow and EDCs resource manager policy details can be found in our previous work [23].

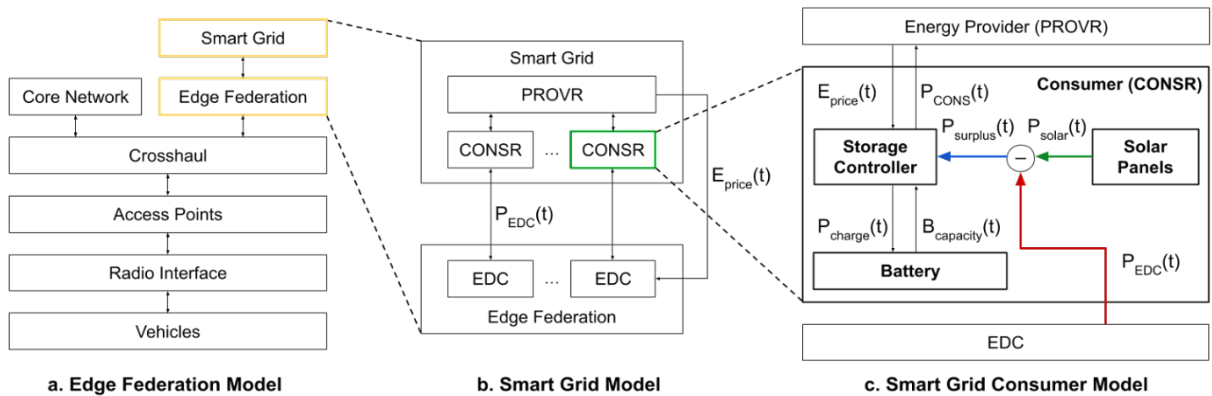
Figure 14: Conceptual Model of the Smart Grid.

This model allows us to explore the coordinated management of power generation, storage, cooling, and computing capabilities for ADAS. We aim to reduce the energy consumption of the federation that delivers the ADAS service by harnessing the heterogeneity of EDCs location, their environmental conditions, and the energy price.

**Performance Evaluation of Harnessing Smart Grid Heterogeneity:**

In this research, we optimized the energy consumption of the federation through workload allocation considering two-phase immersion cooling, user mobility, Smart Grid with solar generation and batteries, and energy prices. We use the same scenario and models for solar generation and energy storage as in our previous work [23], but we do not include the Cloud computing model. In this way, we narrow down the benefits of leveraging Smart Grid heterogeneity to our formal modeling of Edge computing to analyze its sustainability.

We ran a 24-hour scenario in which the ADAS application is served from the Edge infrastructure with a variable demand profile that depends on the vehicles in the coverage range of the system's access points. In this experiment, the ADAS service is delivered to each vehicle from its nearest EDC to isolate the benefits of the Smart Grid. Figure 15.a shows the charging and discharging profile of the batteries of the three EDCs of the federation. The storage controller takes advantage of the off-peak hour to charge the batteries when the energy price is minimal. During peak hours, when the price is higher, the controller prioritizes using stored and solar energy. This increases the demand for grid power during off-peak hours, minimizing its use during peak hours, as shown in Figure 15.b for the federation of three EDCs. Thus, the cost profile of the federated EDCs, presented in Figure 15.c, only shows steep rises when the computation required by the vehicles demands a power consumption that exceeds the solar generation and battery storage capacity. Our results show that harnessing Smart Grid heterogeneity for ADAS can reduce federation energy by 20% and operating costs by 30%, significantly improving the energy and economic sustainability of the Edge infrastructure.

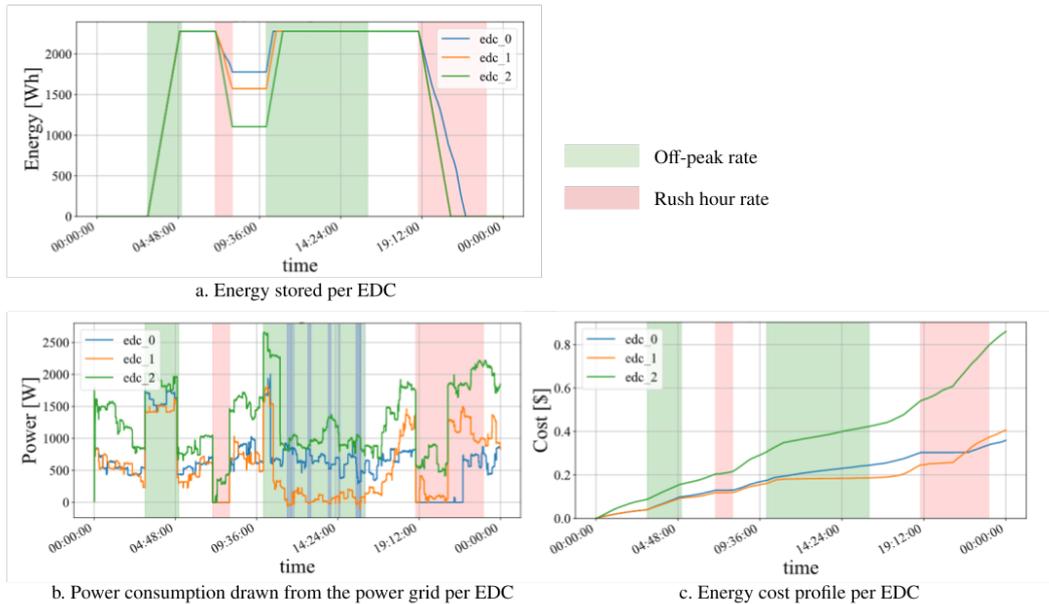

Figure 15: Battery Energy Storage, Grid Power Consumption and Energy Cost.

## VII. Use Case: Fault-Tolerant Cooling Modeling through ML-Based Synthetic Data Generation

Apart from geographic location and renewable production, data science is a major driver of energy efficiency in the Clouds of hyperscale operators. According to DeepMind [24], AI-based solutions reduce cooling energy in Google's data centers by 40%, lowering PUE by 15%. However, the biggest challenge for the industry to include AI-driven optimizations is to get massive amounts of data to train the algorithms properly (especially for DL-based approaches), as it is expensive in terms of costs, time, and resources. Therefore, this technology is reserved for the data owners, as companies usually prefer not to share it to avoid leaking information about their users. This issue is aggravated when ML-driven optimizations must make decisions in anomalous situations since these are sporadic and the amount of data is limited. In this use case, we aim to demonstrate how AI-generated datasets with realistic synthetic scenarios, including on-demand anomalies, can enhance the training of fault-tolerant data center cooling models. In this section, we propose integrating a synthetic scenario generator based on a Generative Adversarial Network (GAN) into our MSOBSE framework to improve the robustness of predictive thermal and humidity models used for the dynamic control of cooling infrastructures.

Predicting temperature in a data center is especially relevant to capture the thermal behavior of the data room. Thermal models automatically manage the setpoint temperature of cooling systems and ensure safe working ranges for the IT equipment. We need reliable and robust predictive thermal models to characterize the impact of cooling optimizations accurately. This allows more aggressive optimizations to be explored to reduce cooling consumption significantly. However, in air-cooled infrastructures, thermal variations are followed by non-linear relative humidity drifts in the data room. Controlling humidity in the data center is critical to avoid irreparable damage to the equipment. The dew point can be reached easily if the humidity is too high. If it is too low, the risk of electrostatic discharges increases. So, to obtain reliable predictive models it is necessary to train them with datasets from scenarios that keep the correlations between temperature and humidity. Moreover, if the models are trained for anomalous situations, they will enable better automatic fault-tolerant control. In this use case, we propose to use a GAN to generate realistic synthetic scenarios that maintain the correlations between variables. GANs allow a high degree of control of the generated scenarios through the latent space, which makes them especially interesting for augmenting biased datasets with on-demand anomalies.

Figure 16 shows the workflow of our system for obtaining a predictive thermal and humidity model and its integration into the control system decision-making process. First, the GAN-based generator provides synthetic scenarios with and without anomalies from real data collected from the data center. Then, real and synthetic datasets are used to train the temperature and humidity model to learn how to make robust predictions in scenarios, also in the event of anomalies. Once trained, our model can be integrated into the MSOBSE framework. Our model, together with the injection of real and synthetic data, will help the control system to explore the impact of optimizations on room temperature and humidity even if anomalies occur. Hence, fault-tolerant decision-making can be improved.

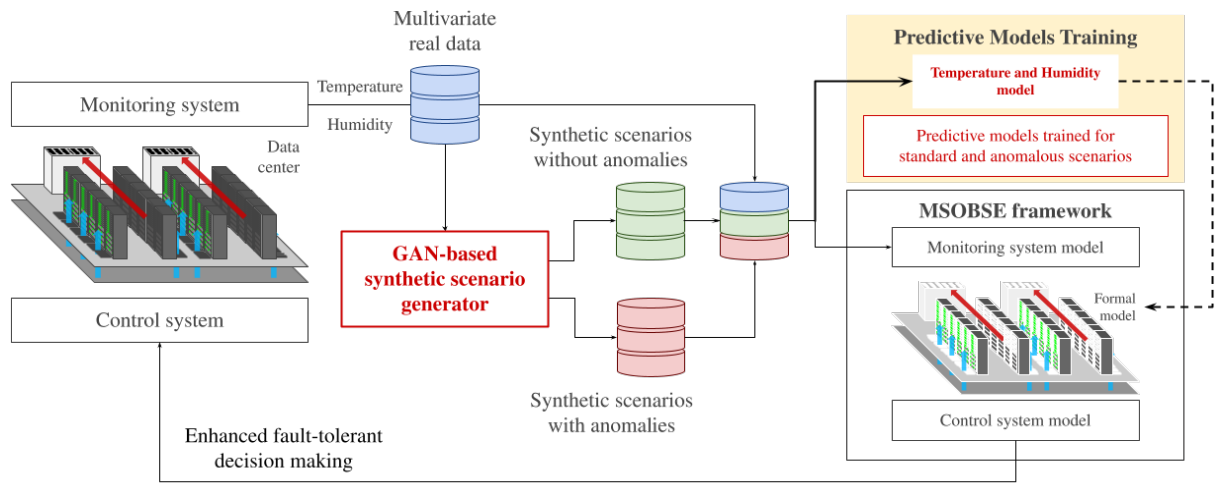

*Figure 16: Workflow of the integrated approach for fault-tolerant cooling modeling.*

**Performance Evaluation of fault-tolerant cooling modeling:**

To evaluate our proposed modeling approach, we have used a dataset of 15,000 temperature and humidity samples from the Adam Data Centers facility in Madrid, Spain. The data samples come from 35 sensors distributed at different heights of the racks in hot and cold aisles, collected every 10 minutes. The dataset also incorporates a categorical variable indicating the unique identifier of the sensor node. These data have been injected into our GAN-based synthetic scenario generator, presented in our previous work [25], to produce synthetic data samples with and without anomalies, as shown in Figure 17.

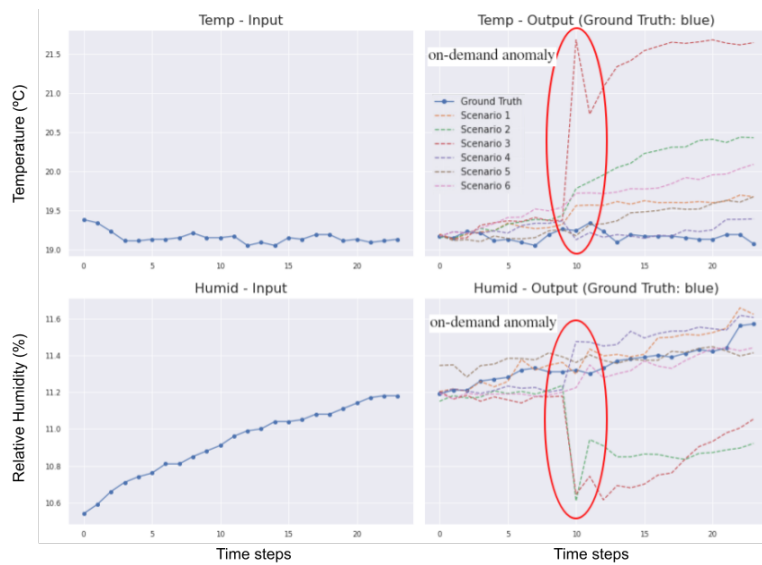

*Figure 17: Synthetic scenario generation for one sensor.*

Figure 17 shows a two-variable time series fed to the GAN and its corresponding output for a single sensor. Each color represents a different 4-hour synthetic scenario for temperature and humidity. In the 10th time step of the generated scenarios, an anomaly has been injected on demand. In scenario 3 we can see a temperature spike that coincides with a drastic reduction in humidity, maintaining the non-linear correlations of the scenario. Next steps of the scenario are consistent with this event providing a realistic evolution (e.g., a server malfunction causes a hot spot, so the fans start working at full speed reducing the temperature, but they cannot extract the heat fast enough and the temperature keeps rising).

To evaluate the training performance with synthetic data, we use a model that obtains temperature and humidity with a prediction window of 10 minutes. The model to be trained is a FeedForward neural network based on Long Short-Term Memory neurons with 2 hidden layers (256 and 32 neurons, respectively) whose 2 output neurons provide the two variables of interest. As in the neural network presented in our previous work [25], we have used an AdaBelief optimizer, a linear output activation, Two Time-Scale Update Rule, and dropout.

First, we trained the model with 15,000 real data samples augmented with 150,000 synthetic samples without injected anomalies. Testing on real data shows that temperature and humidity prediction errors (mean square errors) are reduced by 28.9% and 7.4% respectively, compared to the model trained only with real data. Then, we trained the model with 15,000 real data samples augmented with 150,000 synthetic samples, where 5% have anomalies. Compared to the model trained with real data only, in this case, real data testing shows a reduction in temperature and humidity errors by 27.9% and 17%, respectively. These results show that integrating synthetic data and anomalies can improve model stability and prediction robustness.

## VIII. Conclusions and Future Directions

Edge computing can enable the digitization of society, driving the deployment of critical applications that will improve people's quality of life. But in practice, the implementation of the Edge paradigm is far more disruptive than we anticipated due to the limitations of applying current Cloud-based strategies at the edge of the network. Our main objective is to enhance the sustainable development of future Edge infrastructures, making them environmentally and economically competitive to accelerate their adoption. In this paper, we present our vision for addressing the key challenges of Edge deployment and operation, such as energy efficiency, fault-tolerant automation, and collaborative orchestration. We propose an innovative approach that combines two-phase immersion cooling, formal modeling, simulation and optimization techniques, machine learning, and federated management to drive Edge sustainability.

To demonstrate the benefits of our approach, we present our early steps towards the sustainability of an Edge infrastructure for an Advanced Driver Assistance Systems application. We designed a prototype EDC with a state-of-the-art PUE based on two-phase immersion cooling, exploring its thermal and power behavior. Then, we modeled our prototype data center for the simulation and optimization of its federated operation. Finally, we analyzed the performance of our two-phase immersion-based prototype through the simulation of an Edge federation with Smart Grid-awareness. Our results show that our approach can reduce the Edge federation consumption by 20% and operating costs by 30%, significantly improving the energy and economic sustainability of the Edge infrastructure. We also presented the potential to support automated management by providing synthetic balanced datasets to train more stable and robust models. In this way, optimization algorithms can learn to make decisions under anomalous situations, thus enhancing the fault tolerance of future unattended Edge deployments.

These initial results demonstrate the potential benefits of our vision toward a more sustainable future for Edge computing. In this direction, our future work involves the following research lines.

- Explore the limits of two-phase immersion cooling for higher IT power densities closer to the coolant critical heat flux.

- Analyze new fluid condensation methods to develop free-cooling two-phase immersion systems.

- Study safe workload allocation and consolidation strategies aware of the heat exchange capacity limits of two-phase immersion coolants.

- Improve fault-tolerant operation of Edge federations to increase security against anomalies and vulnerabilities.

- Design sustainable dimensioning and deployment of Edge federations for new critical applications according to their dynamic resource requirements and demand profiles.

- Provide novel energy optimizations for the collaboration of geographically distributed Edge federations and Clouds leveraging their heterogeneity.

- Develop collaborative strategies in competitive multi-tenant environments for critical applications.

## Acknowledgements

We would like to thank 3M, Adam Data Centers, ImesAPI, and Tychetools for their support of this research.

This research is partially supported by the HiPEAC6 Network, financed by the European Union's Horizon2020 research and innovation programme, and by the University of Melbourne. This work has also been supported by the Centre for the Development of Industrial Technology (CDTI) and State R&D Program Oriented to the Challenges of the Society (Retos Colaboración 2017) under contracts IDI-20171194 and RTC-2017-6090-3 from the Spanish Ministry of Science and Innovation, and by the Spanish State Research Agency (Agencia Estatal de Investigación) under grant PID2019-110866RB-I00/AEI/10.13039/501100011033.